# Spin Fluctuation Induced Linear Magnetoresistance in Ultrathin Superconducting FeSe Films

Qingyan Wang[1]*, Wenhao Zhang[2]*, Weiwei Chen[3]*, Ying Xing[1,6], Yi Sun[1], Ziqiao Wang[1], Jia-Wei Mei[4], Zhengfei Wang[3†], Lili Wang[2,5†], Xu-Cun Ma[2,5], Feng Liu[4,5†], Qi-Kun Xue[2,5], Jian Wang[1,5†]

[1]*International Center for Quantum Materials, School of Physics, Peking University, Beijing 100871, China*

[2]*State Key Laboratory of Low-Dimensional Quantum Physics, Department of Physics, Tsinghua University, Beijing 100084, China*

[3]*Hefei National Laboratory for Physical Sciences at the Microscale, CAS Key Laboratory of Strongly-Coupled Quantum Matter Physics, University of Science and Technology of China, Hefei, Anhui 230026, China*

[4]*Department of Materials Science & Engineering, University of Utah*

[5]*Collaborative Innovation Center of Quantum Matter, 100084 Beijing, China*

[6]*Beijing Key Laboratory of Optical Detection Technology for Oil and Gas, China University of Petroleum, Beijing 102249, China*

*Qingyan Wang, Wenhao Zhang, Weiwei Chen contributed equally to this work.
†Corresponding authors:
zfwang15@ustc.edu.cn;
liliwang@mail.tsinghua.edu.cn ;
fliu@eng.utah.edu;
jianwangphysics@pku.edu.cn

**Abstract**
The discovery of high-temperature superconductivity in FeSe/STO has trigged great research interest to reveal a range of exotic physical phenomena in this novel material. Here we present a temperature dependent magnetotransport measurement for ultrathin FeSe/STO films with different thickness and protection layers. Remarkably, a surprising linear magnetoresistance (LMR) is observed around the superconducting transition temperatures but absent otherwise. The experimental LMR can be reproduced by magnetotransport calculations based on a model of magnetic field dependent disorder induced by spin fluctuation. Thus, the observed LMR in coexistence with superconductivity provides the first magnetotransport signature for spin fluctuation around the superconducting transition region in ultrathin FeSe/STO films.

**Introduction**

Recently, FeSe has attracted extensive attention because of its fascinating superconductivity and intriguing magnetic properties [1-14]. The superconducting transition temperature ($T_c$) of bulk FeSe is ~8K [1], but a high $T_c$ of ~40-100K has been measured by different experimental techniques in one unit-cell (UC) FeSe grown on $SrTiO_3$(001) (STO) substrate [2-14]. The enhanced $T_c$ in 2D is generally believed to be associated with the interface effects, while the underlying mechanisms for superconductivity are still not fully clear. Both conventional mechanism of electron-phonon coupling [15-17] and other contemporary pairing mechanisms [18-21] have been considered. For the latter, the relationship between superconducting and magnetic properties is of special interest. For bulk iron-based pnictides superconductors, experiments have observed coexistence of superconductivity with antiferromagnetic (AFM) ordering [22], as well as spin fluctuations around superconducting temperatures by inelastic neutron scattering [23]. For 2D FeSe/STO, combining the theoretical calculation and angle resolved photoemission spectroscopy (ARPES), a checkerboard AFM has recently been identified [24]. Also, spin fluctuation has been proposed to play a significant role in the high $T_c$ mechanism of 2D FeSe/STO [25]. However, direct measurement of spin order and spin fluctuation in FeSe/STO remains too difficult.

Linear magnetoresistance (LMR) is an intriguing phenomenon in the condensed-matter physics, since MR is supposed to be quadratic in conventional materials. LMR has been observed in various systems such as nonmagnetic narrow-band semiconductors $Ag_{2+\delta}Se$ and $Ag_{2+\delta}Te$ [26], InSb [27], epitaxial graphene [28], topological insulators [29], and recent Dirac and Weyl semimetals [30-34]. There are two prevailing models to characterize the LMR, one is the classical model by Parish and Littlewood [35] and the other is the quantum model by Abrikosov [36, 37]. The classical model is proposed for inhomogeneous conducting materials where LMR is linked with electrical disorder, while the quantum model is proposed for gapless materials in quantum limit where LMR is linked with linear Dirac band. Therefore, the MR provides an effective transport measurement for the intrinsic electronic structures of a material. Besides the traditional structure disorders, such as vacancy, adatom and dislocation, the spin fluctuation can be considered as a special type of spin disorder. It's well known that spin moment is very sensitive to the magnetic field, so that exotic MR features are also expected as an indirect signature to detect the spin fluctuation.

In this Letter, we report an *ex situ* magnetotransport study of ultrathin FeSe/STO films. A robust LMR is observed around the superconducting transition temperatures regardless of film thickness and protection layers. Furthermore, magnetotransport calculations have been performed to reproduce the experimentally observed LMR by including a magnetic field dependent disorder induced by spin fluctuation. Our results demonstrate the coexistence of LMR and superconductivity, and provide a defining magnetotransport signature for spin fluctuation in ultrathin FeSe/STO films.

**Experimental**

Four FeSe films with varied thickness (1 UC, 2 UC, two 3 UC samples labeled as S1 and S2) were grown on insulating STO substrates by molecular beam epitaxy (MBE) as described in previous studies [2]. The FeSe films become superconducting after proper annealing in ultra-high vacuum (UHV) [38, 39]. Prior to be transferred out of UHV system for *ex situ* transport measurement, the films were capped with protection layers to avoid contamination and oxidation in atmosphere. For 2 UC and 3 UC FeSe films, 20 nm-thick amorphous Si was directly deposited as the protection layer. Since 1 UC FeSe film is more difficult to survive in atmosphere, it is protected by 10 UC epitaxial non-superconducting FeTe protection layers and additional 20 nm-thick amorphous Si layer on top.

**Results and discussion**

Figure 1 presents the temperature dependence of resistance for the four samples measured in our experiment, where the resistance is scaled by the base value at 100 K. The four films exhibit superconducting transition with varied $T_c$. To be specific, $T_c$ is 33 K, 17 K, 21 K and 25 K for the 1 UC, 2 UC, 3 UC S1 and S2 sample, respectively. Here, $T_c$ is defined as the intersection between the linear extrapolation of the normal state resistance and the superconducting transition regime. According to previous studies, the first FeSe layer on STO is superconducting [2, 39], and $T_c$ varies with carrier doping level therein [38, 39]. Since the annealing condition for each sample is almost the same, the variation in $T_c$ here may be attributed to the different capping layers, i.e. single crystalline FeTe layer or amorphous Si layer. In the latter case, Si atoms may mix into FeSe films and act as disorders, which would induce scattering and in turn degrade the superconductivity in the first FeSe layer. It turns out that an insulating behavior ($dR/dT < 0$) appears before superconducting transition in both 2 UC and 3 UC FeSe samples that are capped with amorphous Si layer.

We observe an unusual positive LMR at temperatures around the superconducting transition region. Figure 2(a) shows transverse MR ratio as a function of perpendicular magnetic field for the 3 UC S1 sample with $T_c \sim 21$ K. Here, the MR ratio is defined as MR($B$) = [$R_{xx}(B) - R_{xx}(0)$]/$R_{xx}(0)$, with $R_{xx}(B)$ and $R_{xx}(0)$ denoting the resistance by standard four-probe measurement under applied magnetic field $B$ and zero magnetic field, respectively. Displayed in the inset is the resistance vs. temperature curve at zero magnetic field with marked temperature points at which we measured the MR by sweeping magnetic fields. In Fig. 2(b) and 2(c), the corresponding power-law fitting (MR $\sim B^n$) and the first-order derivative of MR (dMR/d$B$ versus $B$) are demonstrated, respectively. At temperatures from 15 K to 22 K, consistent observation of the power exponent n $\sim$ 1.04-1.10 (Fig. 2(b)) and constant dMR/d$B$ above a critical magnetic field $B^*$ (Fig. 2(c)) indicate an extraordinary linear response of MR in this regime. Here, $B^*$ is defined as the intercept point of the two straight linear-fitting lines. With increasing temperature, the linear response of MR shrinks gradually and parabolic response emerges. At temperatures of 28 K and 30 K, the power-law fittings yield n $\sim$ 1.72 and $\sim$ 1.87 (Fig. 2(b)), respectively, indicative of parabolic MR. The field dependence of LMR can be clearly discerned from dMR/d$B$ versus $B$ curves shown in Fig. 2(c). Take the data measured at 15 K as an example, dMR/d$B$ develops almost linearly at low magnetic fields from $B = 0$, implying a quadratic magnetoresistive behavior. Above a crossover magnetic field $B^* \sim 0.6$ T, dMR/d$B$ tends to be a constant, indicating a positive LMR behavior. Therefore, the relation of MR with $B$ can be described as a crossover behavior from $B^2$ at low magnetic field to $B$ at high magnetic field. This crossover behavior from the semiclassical regime to the LMR regime has also been reported in single crystal iron chalcogenide $Fe_{1+y}Te_{0.6}Se_{0.4}$ [40].

In that case, LMR has been obtained at high magnetic fields in a wide temperature range above $T_c$, which is different from our observation.

In order to further explore the dependence of LMR on the direction of the applied magnetic field, we perform the magnetotransport study on another 3 UC FeSe film (S2) with $T_c \sim 25$ K under both perpendicular and parallel magnetic fields, as shown in Fig. 3(a) and 3(b), respectively. For this 3 UC FeSe film under perpendicular field at 8 K and 10 K, superconducting dip is observed at low magnetic field, while the LMR behavior appears at high field. At temperatures from 10 K to 20 K, the superconducting dip disappears, while the LMR extends to low fields and does not saturate at magnetic fields as high as 15 T. Above 20 K, as indicated by the power-law fitting (MR $\sim B^n$) shown in Fig. 3(c), the magnetic field dependence of MR evolves from the linear (n $\sim$ 1) to quadratic form (n $\sim$ 2). The LMR temperature range is from 8 K to 20 K, which is located around the superconducting transition region. We point out that the temperature dependence of the power exponent n is a universal phenomenon for FeSe films under perpendicular magnetic field. That is, it is close to 1 (the LMR) around superconducting transition region, but increases monotonically and approaches to 2 at higher temperatures. Furthermore, the LMR behavior also exists in parallel field but in smaller temperature range [the curve at 8 K in Fig. 3(b)]. Therefore, the LMR is more apparent under perpendicular field for FeSe film. To better clarify the relation between the LMR and the film thickness, we further investigate the magnetic response of the 1 UC and 2 UC FeSe films under perpendicular field, as shown in Fig. 3(e) and 3(f), respectively. For 1 UC FeSe film with $T_c \sim 33$ K, the MR decreases gradually with increasing temperatures, similar to the situations in 3 UC films. At low temperatures below 20 K, the superconducting transition dominates at low magnetic field, while the LMR is obtained at high magnetic field. Similar LMR behavior has also been observed in 2 UC FeSe film. In addition, same as the observation in 3 UC samples, the LMR region in 1 UC and 2 UC samples is closely related to the superconducting transition region too. It is noticed that the 2 UC sample here is not of the highest quality since the residual resistance below $T_c$ is large. Furthermore, the protection layer of the 1 UC film is 10 UC FeTe and 20-nm thick amorphous Si layer, while it is only amorphous Si capping layer for 2 UC and 3 UC FeSe films. Based on the data shown in this paper, we conclude that the existence of the LMR behavior is irrelevant to the ultrathin FeSe film thickness and the protection layer.

One striking feature of our experiment is that the LMR in all samples can only be observed around the superconducting transition region. Such a special temperature dependence of LMR cannot be well understood by original LMR models [35-37]. To identify the underlying physics of the LMR in FeSe/STO, first-principles calculations for 1UC FeSe with checkerboard AFM and spin orbital coupling (SOC) are carried out in the framework of GGA with PBE functional using the VASP package. This spin configuration has been indirectly identified as the ground state of FeSe/STO recently [24]. A tight-binding (TB) Hamiltonian of FeSe is obtained from the Wannier fitting of first-principles bands, and the magnetic field is introduced by the standard Peierls substitution [41]. Using the TB Hamiltonian, a FeSe nanoribbon of 60 unit-cell width is constructed and the two-terminal magnetotransport properties are calculated using the Landauer Buttiker formula. To suppress the inter-edge coupling, the disorder is restricted within 40 unit cells away from the edge. The disorder and magnetic field are only applied to the center scattering region with a length of 100 unit cells in the transport calculation (Fig.4(a) and (b)), and the conductance is averaged over 40 configurations of disorder for statistical convergence. In order to reproduce the experimental results, we assume a magnetic field (B) dependent Anderson-typed random on-site disorder [42] as $\varepsilon_i=w_i \cdot s \cdot \ln(\alpha \cdot B+1)$, accounting for the local spin fluctuation of Fe atoms in the superconducting transition region. Here, $s=\pm1$ for spin-up and spin-down component, $w_i$ is a Gaussian distribution with a standard deviation of w, and $\alpha$ is a fitting parameter representing the strength of magnetic field dependence. This assumption is physically reasonable, since the spin fluctuation is largest as the system going through magnetic to superconducting phase. Also, a large spin fluctuation has already been detected by inelastic neutron scattering in this region for bulk FeSe [23]. The theoretical MR results with $\alpha=16$ are shown in Fig. 4(c), which are in qualitative agreement with the experimental results (Fig. 3(e)). In addition, we found that the overall shape of theoretical MR curve depends sensitively on the parameter $\alpha$, indicating a relatively strong magnetic field dependence. Therefore, it is reasonable to expect different $\alpha$ values for FeSe films with different thickness and protection layers. To further support this conclusion, we have reproduced the experimental LMR in Fig. 2(a) by setting $\alpha=1$, as shown in Fig. 4(d). Again, the qualitative agreement is very good.

In the above calculations, the Fermi-level is set in the bulk states. The observed LMR is induced by a form of spin fluctuation, which has no relation with the topological Dirac edge state in FeSe [24]. If the Fermi-level is set within the SOC gap in our calculations, the MR would be near zero. The different magnetic field response for bulk and topological Dirac edge states can be easily understood. Since spin-up and spin-down bands are approximately decoupled in FeSe [24], the spin-flip backscattering is forbidden for the topological Dirac edge states, so that their conductance cannot be suppressed by the disorder. However, the bulk states have backscattering channels even for the same spin component, so that their conductance is very sensitive to the disorder. Consequently, our observed LMR in FeSe films has a dramatically different physical origin as compared to the LMR in HgTe quantum well, which is induced by topological Dirac edge state with non-magnetic field dependent disorder [43].

Lastly, to further clarify the correlation between MR and spin fluctuation in FeSe thin films, we plot MR as a function of the temperature for the fixed magnetic field. As shown in Fig.5(a), MR grows faster when the temperature is below $T_c$ for all the samples, demonstrating the characterized features in the superconducting transition region. This phenomenon is the same to the dynamic spin correlation function measured for bulk FeSe in the same region (Fig. 4(a) in Ref. 23), which is used to show spin fluctuation. Therefore, MR and spin fluctuation are directly linked together in this region. Also, the MR as a function of $T_c$ is plotted in Fig. 5(b). One can see that the higher $T_c$ sample has a larger MR, indicating a larger spin fluctuation to quench down the magnetic moment in entering the superconducting phase. This is consistent with our general understanding about the correlation between AFM and superconductivity [44], and also with our theoretical modeling where a larger value of parameter $\alpha$, which represents the intensity of spin fluctuation in response to magnetic field, is used for the higher $T_c$ sample.

In summary, we have performed a magnetotransport measurement for ultrathin FeSe/STO films and discovered an unusual LMR around the superconducting transition region. Furthermore, a spin fluctuation mechanism is proposed to explain the experimental observation. Our results demonstrate the coexistence of superconductivity and LMR, identifying indirectly a magnetotransport signature for spin fluctuation in ultrathin FeSe/STO films. We envision our findings may inspire future studies to further investigate the interplay between magnetism and exotic magnetotransport properties in 2D superconducting materials.

**Acknowledgement**

We thank Xingao Gong for discussions. The experimental work was supported by the National Basic Research Program of China (Grant No. 2013CB934600 and and 2015CB921000) and National Science funding of China (Grants No. 11574174), the Open Project Program of the Pulsed High Magnetic Field Facility (Grant No. PHMFF2015002) at the Huazhong University of Science and Technology, the Open Research Fund Program of the State Key Laboratory of Low-Dimensional Quantum Physics, Tsinghua University under Grant No. KF201501, and the Science Foundation of China University of Petroleum, Beijing (2462017YJRC012). The theoretical work was supported by the National Natural Science Foundation of China (Grant No. 21603210), the Chinese Youth One Thousand Talents Program, the Fundamental Research Funds for the Central Universities, and the United States Department of Energy Basic Energy Sciences (Grant No. DE-FG02-04ER46148). We also thank Supercomputing Center at USTC, CHPC at University of Utah and NERSC for providing the computing resources.

Figure caption

FIG. 1. (Color online) Temperature dependence of the resistance scaled by the values at 100 K under zero magnetic field for 1 UC, 2 UC and two 3 UC FeSe (S1 and S2) films grown on STO substrates. The values mark the $T_c$ of the corresponding FeSe films.

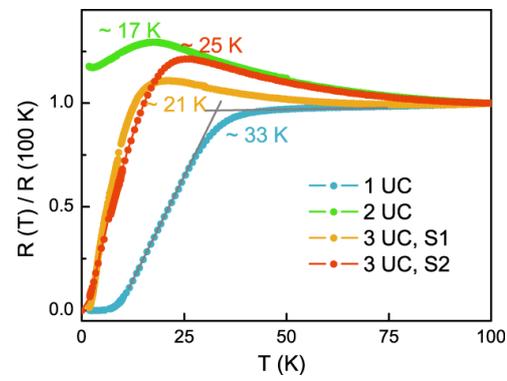

FIG. 2. (Color online) (a) Magnetic field dependence of MR for one 3 UC FeSe film (S1) at different temperatures. The inset labels the corresponding temperature data points at which we measure the MR by sweeping magnetic fields. (b) A double-logarithmic plot of MR versus magnetic fields at different temperatures. The lines show the power-law fitting (MR ~ $B^n$) of each curve in double-logarithmic coordinates above 3 T. The power exponent n is shown at the right column. (c) The field derivative of MR at different temperatures. The linear-fitting solid lines show the crossover from semiclassical regime to the linear response region.

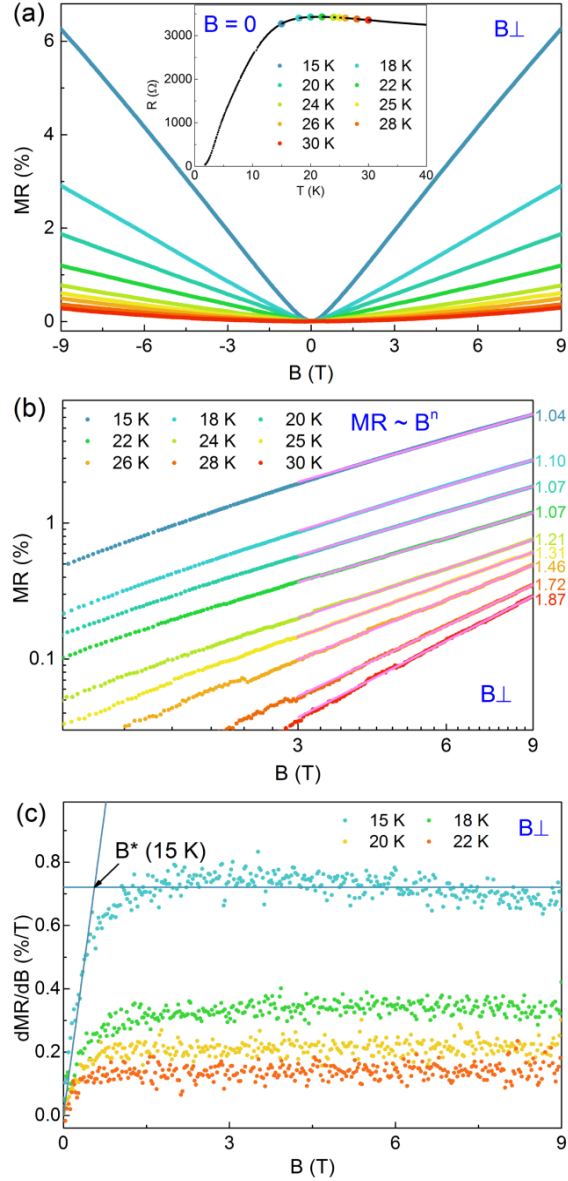

FIG. 3. (Color online) (a) and (b) MR as function of magnetic fields applied in perpendicular and parallel direction for another 3 UC FeSe film (S2). Insets: the temperature data points at which we measure the MR. (c) and (d) demonstrate the MR as function of magnetic fields at different temperatures in double-logarithmic plot in perpendicular and parallel direction, respectively. (e) and (f) demonstrate the magnetic field dependence of MR for 1 UC FeSe film and 2 UC FeSe film, respectively. Insets: the temperature data points at which we measure the MR.

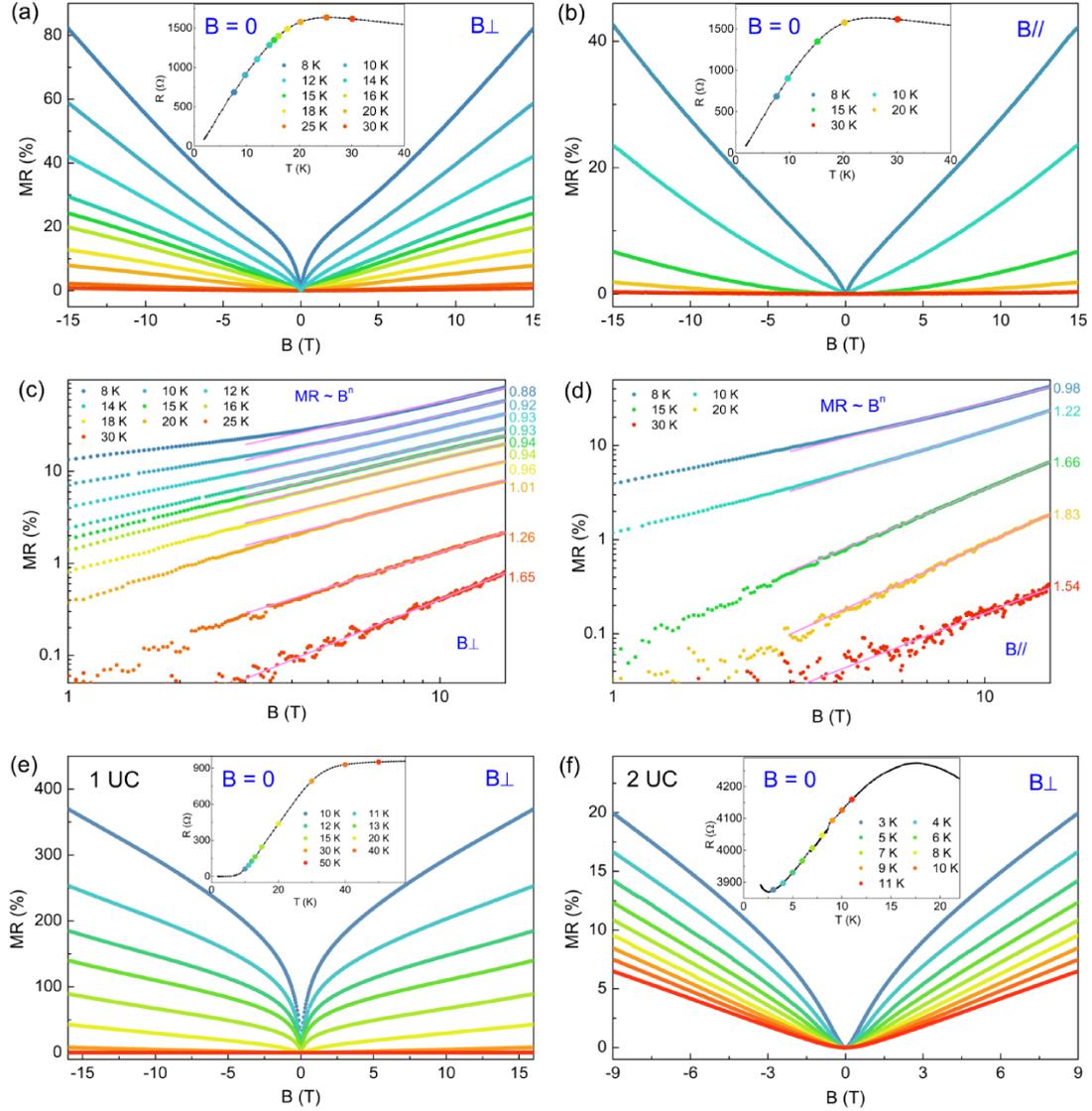

FIG. 4. (Color online) (a) Device setup of magnetotransport calculations. (b) Zoom-in view of the spin fluctuation induced disorder. (c) and (d) MR of FeSe with fitting parameter $\alpha=16$ and $\alpha=1$, respectively.

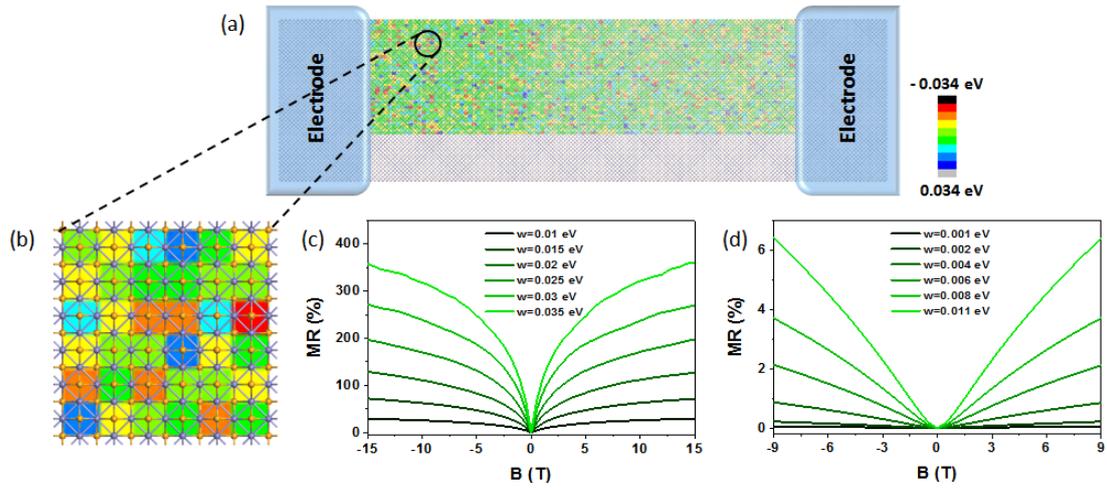

FIG. 5. (Color online) (a) Temperature dependence of MR at 9 T for different samples. (b) MR at 9 T as a function of $T_c$ for different samples measured at different temperatures.

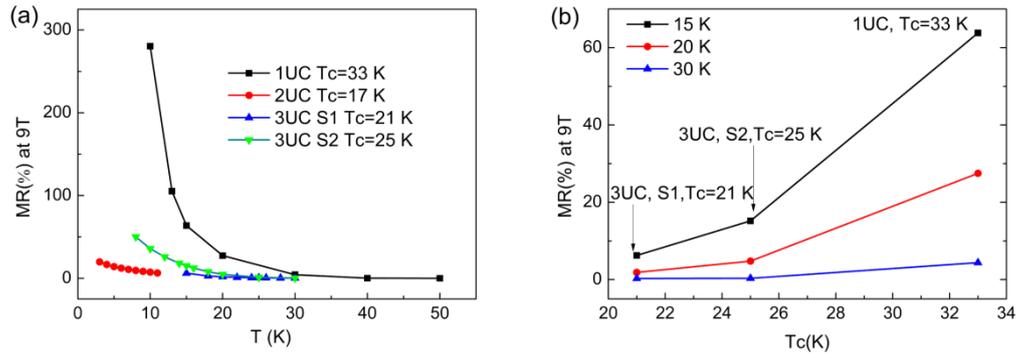